\documentclass[12pt]{article}
\usepackage{epsfig}
\usepackage{latexsym}
\begin{document}
\begin{center}

{ \bf KINETIC ENERGY APPROACH TO DISSOLVING AXISYMMETRIC
MULTIPHASE PLUMES}

\vspace{1cm}

Kristian Etienne Einarsrud\footnote{E-mail: k.e.einarsrud@gmail.com}
 and Iver Brevik\footnote{E-mail: iver.h.brevik@ntnu.no}

 \bigskip

 Department of Energy and Process Engineering, Norwegian
 University of Science and Technology, N-7491 Trondheim, Norway

 \bigskip

 \bigskip

 \today
\end{center}

\begin{abstract}
A phenomenological kinetic energy theory of buoyant multiphase
plumes is constructed, being general enough to incorporate the
dissolution of the dispersive phase. We consider an axisymmetric
plume, and model the dissolution by means of the Ranz-Marshall
equation in which there occurs a mass transfer coefficient
dependent on the plume properties. Our kinetic energy approach is
moreover generalized so as to take into account variable slip
velocities.

The theoretical model is compared with various experiments, and
satisfactory agreement is found. One central ingredient in the
model is the turbulent correlation parameter, called $I$, playing
a role analogous to the conventional entrainment coefficient
$\alpha$ in the more traditional plume theories. We  use
experimental data to suggest a relationship between $I$, the
initial gas flux at the source, and the depth of the gas release.
This relationship is used to make predictions for five distinct
case studies.

Comparison with various experimental data shows that the kinetic
energy approach built upon use of the parameter $I$ in
practice has the order of predictive power as the conventional
entrainment-coefficient models. Moreover, an advantage of the
present model is that its predictions are very quickly worked out
numerically.

Finally, we give a sensitivity analysis of the kinetic energy
approach. It turns out that the model is relatively stable with
respect to most of the input parameters. It is shown that the
dissolution is of little influence on the dynamics of the
dispersed phase as long as the dissolution is moderate.

\bigskip


\end{abstract}

\section{Introduction}

The motivation for studying bubble plumes is evident, from the
fact that these plumes are encountered in a variety of engineering
problems. To mention a few, plumes have been used to damp sea
waves in harbours (pneumatic breakwaters), to prevent surface ice
formation in harbours, to mix stratified fluid layers, to re-aerate
lakes, and finally to protect installations from shock waves
produced by underwater explosions (cf., for instance, Bhaumik,
2005).

With increasing subsea activities plumes have acquired increased
importance from a risk assessment point of view. It becomes
imperative to obtain knowledge about the implications of a
rupture, or even a breakage, of a subsea pipeline. Thus in case of
an underwater blowout of inflammable gas, one wishes answers to
the following questions:

$\bullet$ What is the concentration of gas at the free surface?

$\bullet$ What is the extension of the area covered by gas at the
free surface?

$\bullet$ What is the rising time of the gas?

The aim of the present work is to provide a theoretical framework
from which answers to these questions can be given, in a quick and
reasonably accurate way. Our approach is based on the kinetic
energy method, proposed by one of the present authors some years
ago in the case of air-bubble plumes (Brevik 1977, 1996,1999). The
essential generalization of the present paper is that we allow for
dissolution of the gas. Therewith a realistic answer to the first
question above becomes in principle obtainable.  Especially when
the plume is arising from large depths, the gas dissolution is
expected to be an important factor. 

Figure 1 sketches a plume coming from a gas release at depth $D$.
It is useful to divide the plume into three different zones: The
zone of flow establishment (ZFE) is characterized by high gas
concentration and flow conditions changing rapidly with the height
$z$ above the real source (we thus do not introduce a virtual
source being displaced some distance downward from the real
source). Then, there is a surface zone, characterized by an
outward radial flow of entrained fluid, yielding a rapid increase
of the effective width of the plume. The third zone, the one of
main interest here, is the established flow zone characterized by
the self-similarity property, implying that the flow has
established some sort of self-governing  equilibrium  (Socolofsky
et al., 2002). This means that the time averaged cross-sectional
profiles of the plume parameters maintain a near-Gaussian form.
The plume parameters are fully determined by a centerline value
and a width, both being a function of the height $z$. The Gaussian
form of the plume parameters has been verified in several
experiments; cf., for instance, Kubasch (2001).

\begin{figure}[h!t!bp]
\makebox[\textwidth][c]{
		\includegraphics[width=10cm]{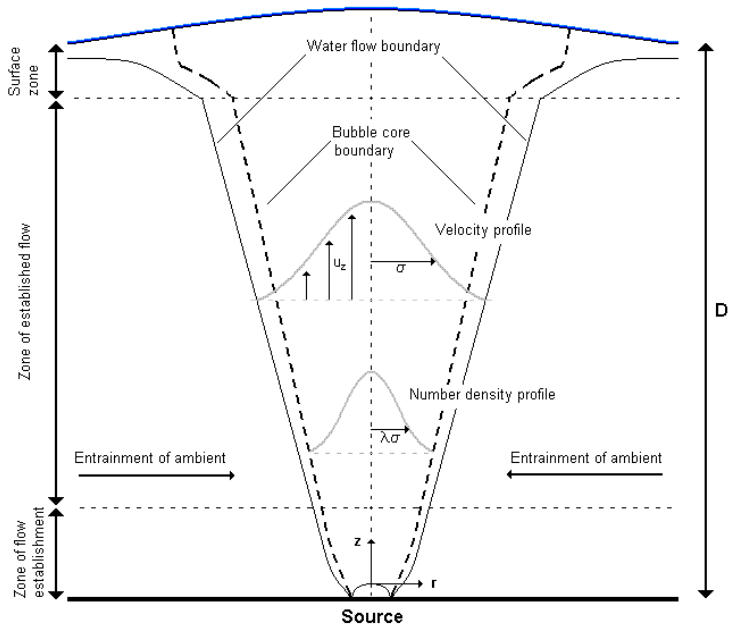}
		}
	\caption{Sketch of rising plume in an unstratified environment originating from a depth $D$ below the (still water) surface. The plume consists of two boundaries, one containing the bubbles intermixed with entrained water (bubbly core, sketched with dashed lines) having a characteristic width $\lambda \sigma$, and one denoting the outer boundary of upwards movement of water (water flow boundary, solid line) having a characteristic width $\sigma$. The effect a radial jet near the surface is sketched by an rapid increase in plume the plume width within the surface zone. Figure inspired by similar graphics found in Kubasch (2001)}
	\begin{center}
	\label{fig:plume}
	\end{center}
\end{figure}

As shown in Fig. \ref{fig:plume} the radius of the plume is $\sigma$, describing
the water boundary, and the centerline vertical velocity $u_c$.
Furthermore, the dispersive phase is bounded by a characteristic
radius $\lambda \sigma$, where $\lambda <1$. The parameter
$\lambda$ is introduced in order to account for the fact that the
dispersed phase forms a "core", whereas the entrained fluid forms
a wider conical structure. $\lambda$ is thus the relative width of
the inner core compared to the outer water flow boundary.

In the present paper we will be concerned with the zone of
established flow only, as this zone is the dominant one in cases
of moderate or large depths. We will ignore the phenomenon of
detrainment (or peeling), which may occur in the case of
stratification, or if the density of entrained fluid is
significantly increased due to dissolution of the dispersive phase
(i.e., the gas bubbles).

It is useful to distinguish between {\it single-} and {\it
multiphase} plumes. The main difference being in the discrete
nature of the buoyant dispersive phase. In a single phase plume
the buoyancy is well mixed with the bulk fluid and the advection
of buoyancy is controlled by the motion of fluid alone. For the
multiphase plume the dispersed phase itself supplies the buoyancy
to the plume and the distribution in the plume is controlled both
by its own dynamics and by the motion of the bulk plume fluid.
This distinction is especially important in the case of
stratification, and in the case of cross-flows. If cross-flows are
present a separation between the dispersed and entrained phases
can occur, yielding a behaviour qualitatively different from the
characteristic plume behaviour. An extensive experimental study of
multiphase plumes has been carried out by Socolofsky (2001).

A great deal of effort has been laid down to describe the
behaviour of multiphase plumes. The main focus of modelling has
been on integral models, considering the integral of relevant
plume parameters over appropriate control volumes. The dynamics of
the plume parameters is governed by coupled differential equations
derived from first principles and/or closure models for
turbulence, mass transfer, heat transfer, and entrainment.

According to the entrainment hypothesis, as introduced by Morton
et al. (1956), the rate of entrainment at the edge of the plume is
proportional to some characteristic velocity at that height. The
characteristic velocity is taken to be the centerline velocity
$u_c$ at that height, and the proportionality constant is the
entrainment coefficient $\alpha$. Mathematically,
\begin{equation}
\frac{dQ_w}{dz}=2\pi \sigma \alpha u_c, \label{1}
\end{equation}
where $Q_w$ is the volumetric flux of entrainment water.

Two different strategies are followed when dealing with multiphase
plumes. They depend in turn on the way in which the multiphase
nature of the problem is treated. A single plume model (or
mixed-fluid model) treats the plume as a mixture of dispersed
phase and entrained fluid. If the ambient fluid can be considered
as a stagnant and unstratisfied fluid, one calls this a {\it
simple} plume.

Simple plume models have been considered in various contexts.
Morton et al. (1956) used a simple plume model in their paper
introducing the entrainment hypothesis; Milgram (1983) compared
his experimental results with the values predicted by a simple
plume model, whereas W{\"u}est et al. (1982) used both a simple
plume model with the entrainment hypothesis  to describe the effect
of dissolving plumes. The work of Brevik (1977) is also based on a
simple plume formulation, though without making use of the
entrainment hypothesis.

A different strategy used to deal with the multiphase nature of
the plume is to use the concept of two-fluid models, namely to
treat each phase separately and introduce coupling terms in the
differential equations for each phase. This superimposition of two
plume-like structure leads to a formalism called a {\it
double-plume} model (or two-fluid model). In the past, two-fluid
models have been implemented only together with the entrainment
hypothesis. McDougall (1978) modelled the system as an inner
circular plume containing the gas bubbles and some entrainment
water, and an annular plume containing only water. The formalism
of McDougall was generalized by Asaeda and Imberger (1993), and by
Crounse et al. (2007), simplifying the implementation of density
effects of dissolving bubbles. An extensive comparison between
mixed and two-fluid models based on the entrainment hypothesis is
given by Bhaumik (2005). The advantages of the integral models are
that the governing equations allow insight into the flow dynamics;
they are computationally efficient and produce reasonable results
in many cases. A drawback is  that the integral models lose their
validity as the plume becomes less self-similar (Socolofsky et al,
2002).

In the next section we summarize the key properties of the kinetic energy model. In section 3 we solve the nondimensional governing equations and determine the values of the parameters, especially the correlation parameter $I$, so as to get a reasonably good agreement with existing experiments. Section 4 is devoted to large-scale case studies, of importance for the oil industry. In section 5 we analyze the sensitivity of the introduced parameters.

\section{Model formulation}

The model forming the basis of our present
investigation is the kinetic energy model presented by Brevik and
Killie (1996) for axisymmetric air-bubble plumes, now generalized
so as to account for dissolution of the dissolved phase. Here the
semi-empirical Ranz-Marshall equation (1952) is important for
predicting the gas dissolution. This is moreover similar to the
approach of W{\"u}est et al. (1992); they describe dissolving
plumes using the entrainment hypothesis.  Our model is based on a
mixed-fluid formalism, and intends to describe a simple bubble
plume in the established zone. We shall predict the width, dissolution and the
rise time of a plume originating from the source lying at
depth $D$ ($z=0$).

\subsection{Mass balance}

Consider first the gas dissolution. The rate of mass transferred
from the dispersed phase to the ambient fluid follows from the
Ranz-Marshall equation (1952):
\begin{equation}
\frac{dm_b}{dt}=-4\pi r_b^2K(c_s-c_i), \label{2}
\end{equation}
where $m_b$ is the mass of one spherical bubble with radius $r_b$,
$c_s$ is the solubility of the dispersed phase, $c_i$ is the local
concentration of the dissolved species in the ambient fluid, and
$K$ is the mass transfer coefficient.

Whereas the original Ranz-Marshall equation was derived to
describe the evaporation of pure liquid drops, the equation has
later been successfully applied to describe dissolution of gas
bubbles (W{\"u}est 1992, Crounse 2007, Johansen 2000), and has
become a standard in this area of research. The coefficient $K$
depends on bubble size and flow parameters, as discussed by
W{\"u}est (1992), Zheng and Yapa (2002), and Crounse (2007).

Zheng and Yapa (2002) combine equations from earlier works to
derive a general formula for $K$ for bubbles in water. Assuming
spherical bubbles of radius $r_b$,
\begin{equation}
K = 0.0113\sqrt{\frac{u_s \mathcal{D}}{r_0 + 0.1r_b}} \label{3}
\end{equation}
where $r_0=0.45~$cm, $u_s$ is the bubble slip velocity, and
$\mathcal{D}$ is the molecular diffusivity of gas in the liquid.

As the mass of a bubble is $m_b=(4\pi/3)r_b^3 \rho_g$ with
$\rho_g$ the gas density, we may insert this into (\ref{2}) and
moreover make use of the equation
\begin{equation}
\rho_g(z)= \frac{\rho_{gs}(\breve{D}-z)}{\breve{P}_{atm}} \label{4}
\end{equation}
which follows from the assumption of isothermal expansion. Here
$\rho_{gs}$ is the gas density at the surface, $\breve{P}_{atm}$
is the pressure at the free surface expressed as a head of water
column, and $\breve{D}=\breve{P}_{atm}+D$. This leads
to the following equation for the height dependence of the bubble
radius:
\begin{equation}
\frac{d r_b}{dz} = - \frac{K \breve{P}_{atm} (c_s - c_i)}{u_{zb}\rho_{gs}(\breve{D}-z)} + \frac{r_b}{3(\breve{D}-z)}, \label{5}
\end{equation}
$u_{zb}$ being the vertical velocity of a bubble.

Consider next the mixed-fluid continuity equation by summing the
standard continuity equations for each phase over the two phases.
The density $\rho$ of the mixture is
\begin{equation}
\rho= \alpha_l \rho_l+\alpha_g \rho_g, \label{6}
\end{equation}
where $\alpha_l, \alpha_g$ are the phase fractions of the liquid
and gas phases. Assuming that $\alpha_g \ll \alpha_l$ (or
$\alpha_l \approx 1$), the last term in (\ref{6}) is negligible.
Furthermore, we assume
\begin{equation}
\rho \approx \rho_l \approx \rho_w, \label{7}
\end{equation}
where $\rho_w$ is the ambient fluid density, taken to be constant.
Consequently the (stationary) continuity equation for an axisymmetric plume can be written as
\begin{equation}
\frac{1}{r}\frac{\partial}{\partial r} (r u_r) +  \frac{\partial}{\partial z} u_z = 0 \label{8}
\end{equation}

\subsection{Momentum equation}

Performing the same kind of averaging procedures as in earlier
papers (Brevik and Killie 1996, Brevik and Kluge 1999), now
including gas dissolution, and using the approximation (\ref{7})
meaning that all momentum contributions from the dissolved phase
are neglected apart from the buoyant term, we can write down the
$z$ component of the momentum equation for the fluid mixture. We
assume that the plume is fully turbulent, so that the velocity
field can be Reynolds decomposed into a mean value and a
fluctuating term. Viscous stresses are small and negligible
compared to turbulent stresses. For the averaged Navier-Stokes
equation we obtain, in standard notation,
\begin{equation}
\rho_w[ \bar{u}_r \frac{\partial}{\partial r}\bar{u}_z +\overline{u'_r \frac{\partial}{\partial r}u'_z} + \bar{u}_z \frac{\partial}{\partial z} \bar{u}_z +\overline{u'_z \frac{\partial}{\partial r}u'_z}]= -\frac{\partial}{\partial z}\bar{p} - \rho_w g + \rho_w \alpha_g g \label{9}
\end{equation}
where $g$ is the gravitational acceleration. The mean pressure
$\bar{p}$ can be written as a sum of a dynamic and a hydrostatic
part, $\bar{p}=\bar{p}_d+\rho_w g(D-z)$. Thus the right hand side
of (\ref{9}) simplifies to $\rho_w \alpha_g g$. Introducing the
Reynolds stresses,
\begin{equation}
\tau_{rz}=-\rho_w\overline{u_r'u_z'}, \quad
\tau_{zz}=-\rho_w\overline{u_z'^2}, \label{10}
\end{equation}
and making use of the continuity equation (\ref{8}), we can write
the correlations in (\ref{9}) as
\begin{equation}
\rho_w [\overline{u'_r \frac{\partial}{\partial r}u'_z}  +\overline{u'_z \frac{\partial}{\partial r}u'_z}] = - \frac{\partial}{\partial r}\tau_{rz} - \frac{\partial}{\partial z}\tau_{zz} - \frac{\tau_{rz}}{r} \approx - \frac{1}{r}\frac{\partial}{\partial r} (r \tau_{rz}) \label{11}
 \end{equation}
 where it is assumed that the lateral variations of the Reynolds
 stresses are dominating.

 The momentum equation thus reduces to the form
 \begin{equation}
 	\frac{\partial}{\partial z} \bar{u}_z^2 + \frac{1}{r}\frac{\partial}{\partial r} (r\bar{u}_r\bar{u}_z) = \alpha_g g +  \frac{1}{\rho_w r}\frac{\partial}{\partial r}(r\tau_{rz}). \label{12}
 \end{equation}
 This equation is integrated over a cylindric control volume
 having the boundaries $z=[0, z']$ ($z'$ arbitrary), and
 $r=[0,\infty]$. The boundary conditions are summarized in Table
 \ref{tab:bound}.
 
 \begin{table}[ht!b!p!]
\caption{Boundary conditions for integration of momentum and kinetic energy equations}
\begin{center}
\begin{tabular}{cc}
\hline
Limit & Condition  \\
\hline
$r = 0$ &	$\bar{u}_r = 0$\\
{} &	$r\tau_{rz} = 0$\\
{} &	$\bar{u}_z:$ finite\\
\hline
$r \rightarrow \infty$ &	$\bar{u}_z = 0$\\
{} &	$r\tau_{rz} \rightarrow 0$\\
{} &  $r\bar{u}_r:$ finite\\
\hline
\end{tabular}
\end{center}
\label{tab:bound}
\end{table}

 Let $\rho_n$ be the number density of bubbles. The void fraction
 $\alpha_g$ is related to $\rho_n$ via
 \begin{equation}
 \alpha_g=\rho_n \frac{4\pi}{3}r_b^3. \label{13}
 \end{equation}
 With the given boundary conditions, the integration of (\ref{12})
 leads to
 \begin{equation}
	2\pi\int_0^{\infty} \bar{u}_z^2rdr = 2\pi g \int_0^{z'}\int_0^{\infty}\alpha_g rdzdr = \frac{8\pi^2}{3} g \int_0^{z'} r_b^3\int_0^{\infty}\rho_n rdzdr. \quad \label{14}
 \end{equation}
 In order to advance further, assumptions must be made about the
 form of $\bar{u}_z$ and $\rho_n$. We shall assume Gaussian
 distributions:
 \begin{equation}
	\bar{u}_z (z,r) = u_c(z)e^{-\frac{r^2}{2\sigma^2}} \label{15}
 \end{equation}
 \begin{equation}
\rho_n(z,r) = \rho_{nc}(z)e^{-\frac{r^2}{2\lambda^2\sigma^2}} \label{16}
 \end{equation}
 where $u_c(z)$ and $\rho_{nc}(z)$ are (unspecified) centerlines
 values. The plume grows wider as the free surface is approached,
 mainly due to turbulent diffusion, which corresponds to
 $\sigma=\sigma(z)$.

 Insertion of (\ref{15}) and (\ref{16}) into (\ref{14}) leads to
 the relation
\begin{equation}
 u_c^2\sigma^2 = \frac{8\pi}{3} g\lambda^2 \int_0^{z'} r_b^3\rho_{nc}\sigma^2 dz.
	\label{17}
\end{equation} Upon differentiation,
 \begin{equation}
	\frac{d}{dz}u_c^2\sigma^2 = \frac{8\pi}{3} g\lambda^2 r_b^3\rho_{nc}\sigma^2 = 2 g\lambda^2 V_b\rho_{nc}\sigma^2\label{18}
 \end{equation}
 where $V_b=V_b(z)$ is the volume of a single bubble.

 Assuming that the number of bubbles is constant, in other words
 that there is an equilibrium between bubble coalescence and break
 up, the centerline number density $\rho_{nc}$ can be expressed in
 terms of the number flux density $\dot{N}$ of bubbles. Evaluating
 the basic expression
 \begin{equation}
 \dot{N}=2\pi \int_0^\infty \rho_nu_brdr, \label{19}
 \end{equation}
 we obtain
 \begin{equation}
 \rho_{nc} = \frac{\dot{N}(1+\lambda^2)}{2\pi\lambda^2 \sigma^2 [u_c + u_s (1+\lambda^2)]} \label{20}
 \end{equation}
 Here the factor $u_c+u_s(1+\lambda^2)$ may be interpreted as a
 typical vertical bubble velocity, i.e.
 $u_{zb}=u_c+u_s(1+\lambda^2)$.

 Inserting (\ref{20}) into (\ref{18}) we obtain the final form
 \begin{equation}
		\frac{d}{dz}u_c^2\sigma^2 = \frac{g V_b \dot{N}(1+\lambda^2) }{\pi [u_c + u_s (1+\lambda^2)]}. \label{21}
 \end{equation}
 It is desirable here to check with the known result in the case of no gas
 dissolution. Putting $K=0$ in (\ref{5}) and integrating with
 respect to $z$, we get
 \begin{equation}
\frac{r_b}{r_{bs}} = \Big( \frac{\breve{P}_{atm}}{\breve{D}-z}\Big) ^{\frac{1}{3}}. \label{22}
 \end{equation}
 Cubing this equation we get
 \begin{equation}
 V_b = V_{bs} \frac{\breve{P}_{atm}}{\breve{D}-z}. \label{23}
 \end{equation}
 which after insertion into (\ref{21}) yields
 \begin{equation}
\frac{d}{dz}u_c^2\sigma^2 = \frac{g V_{bs} \dot{N}\breve{P}_{atm}(1+\lambda^2) }{\pi (\breve{D}-z)[u_c + u_s (1+\lambda^2)]} =   \frac{g Q^0 \breve{P}_{atm}(1+\lambda^2) }{\pi (\breve{D}-z)[u_c + u_s (1+\lambda^2)]} \label{24}
 \end{equation}
 where $Q^0=V_{bs}\dot{N}$ agrees with Eq.~(22) in Brevik and
 Killie (1996).

 \subsection{Kinetic energy equation}

 Equations (\ref{5}) and (\ref{21}) contain three unknown
 quantities, $u_c, r_b$ and $\sigma$ (assuming the other
 quantities can be determined from second principles or
 experiments). In order to close the system of equations, a third
 equation is needed. In our approach, this equation will be the
 conservation equation for kinetic energy.

 The starting point is the momentum equation in the form
 (\ref{12}). Taking again into account the continuity equation
 (\ref{8}), we obtain after some manipulations
 \begin{equation}
 \frac{\partial}{\partial z} (\frac{1}{2} \bar{u}_z^3) + \frac{1}{r} \frac{\partial}{\partial r}(\frac{1}{2}r\bar{u}_r\bar{u}_z^2)  + \frac{\bar{u}_z}{r} \frac{\partial}{\partial r}(r \overline{u'_r u'_z}) = \alpha_g \bar{u}_z g. \label{25}
 \end{equation}
 This equation corresponds to (\ref{12}) in the momentum case. We
 integrate (\ref{25}) over the same control volume as before, and
 apply the boundary conditions given in Table 1. Then
 differentiating the equation with respect to $z$ we get the
 integral-differential equation
 \begin{equation}
\frac{d}{dz} \int_0^{\infty} \frac{1}{2} \bar{u}_z^3 rdr = \int_0^{\infty} r \overline{u'_r u'_z} \frac{\partial}{\partial r}\bar{u}_z dr + g\int_0^{\infty} \alpha_g \bar{u}_z rdr. \label{26}
 \end{equation}
 In addition to the models for $u_z$ and $\alpha_g$, a model for
 the turbulent correlation $\overline{u_r'u_z'}$ is needed. As in
 Brevik and Killie (1996) we assume self-preservation:
 \begin{equation}
 \frac{\overline{u_r'u_z'}}{u_c^2}=f(\eta), \label{27}
 \end{equation}
 where $f$ is an unspecified function of the nondimensional
 parameter $\eta=r/\sigma$. making use of this, together with the
 Gaussian forms (\ref{15}) and (\ref{16}), we obtain finally
 \begin{equation}
\frac{d}{dz}u_c^3\sigma^2 = \frac{3}{\pi}\frac{g V_b u_c \dot{N}}{u_c + u_s(1+\lambda^2)} - u_c^3\sigma I \label{28}
 \end{equation}
 where
 \begin{equation}
	I = 6\int_0^{\infty} f\eta^2 e^{-\frac{1}{2}\eta^2}d\eta \label{29}
 \end{equation}
 is an unspecified constant to be evaluated from experiments.

 It is easily shown that (\ref{28}) is consistent with the results
 of Brevik and Killie (1996) in the limit $K\rightarrow 0$.

 Equations (\ref{5}), (\ref{21}) and (\ref{28}) form a closed set
 for the unknowns $r_b, u_z$ and $\sigma$. For given initial
 conditions, if the parameters $u_s, \lambda$ and $I$ are
 determined from experiments, the equations determine the unknown
 quantities at an arbitrary height within the zone of established
 flow.

 \section{Method of solution}

 \subsection{Nondimensional formulation}

 For computational purposes it is convenient to express the
 equation in nondimensional form. We define the parameters
 \[
z^*=\frac{z}{\breve{D}}, \quad r_b^*=\frac{r_b}{\breve{D}}, \quad
K^*=\frac{K}{w_0}, \]
\begin{equation}
u_s^*=\frac{u_s}{w_0}, \quad u_c^*=\frac{u_c}{w_0}, \quad
\gamma^*=\frac{\breve{P}_{atm}(c_s-c_i)}{\rho_{gs}\breve{D}},
\label{30}
\end{equation}
where $w_0$ is a reference velocity defined in section 3.3. Equation (\ref{5}) can now be written
\begin{equation}
\frac{d r_b^*}{d z^*} = -\frac{\gamma^* K^*}{(u_c^*+(1+\lambda^2)u_s^*)(1-z^*)} + \frac{r_b^*}{3(1-z^*)} \label{31}
\end{equation}
Further, (\ref{21}) and (\ref{28}) become
\begin{equation}
\frac{d}{dz^*}(u_c^*\sigma^*)^2 = \frac{(r_b^*)^3(1+\lambda^2)}{u_c^*+u_s^*(1+\lambda^2)} \label{32}
\end{equation}
\begin{equation}
	\frac{d}{dz^*}\left[u_c^*(u_c^*\sigma^*)^2\right] = \frac{(r_b^*)^3 u_c^*}{u_c^*+u_s^*(1+\lambda^2)} - (u_c^*)^3\sigma^*I^* \label{33}
\end{equation}
with
\begin{equation}
	I^* = \frac{1}{2}\sqrt{\frac{3 w_0^3}{\breve{D}^2 g \dot{N}}}I. \label{34}
\end{equation}
In (\ref{33}), we have introduced
\begin{equation}
		\sigma^* = \sqrt{\frac{3w_0^3}{4\breve{D}^4g\dot{N}}}\sigma \label{35}
\end{equation}
Further introducing
\begin{equation}
\kappa^* = (\sigma^* u_c^*)^2 = \frac{3\sigma^2 u_c^2 w_0}{4\breve{D}^4g\dot{N}} \label{36}
\end{equation}
we may after some manipulations replace (\ref{21}) and (\ref{28})
with
\begin{equation}
\frac{d}{dz^*}u_c^* = \frac{u_c^* (r_b^*)^3 (2-\lambda^2) }{\kappa^* (u_c^*+u_s^*(1+\lambda^2))} - \frac{(u_c^*)^2} {\sqrt{\kappa^*}}I^* \label{37}
\end{equation}
\begin{equation}
\frac{d}{dz^*}\kappa^* = \frac{(r_b^*)^3(1+\lambda^2)}{u_c^*+u_s^*(1+\lambda^2)}. \label{38}
\end{equation}
The three nondimensional equations (\ref{31}), (\ref{37}) and
(\ref{38}) are solved iteratively using the ODE45 solver in
Matlab. This solver is a Runge-Kutta method for numerical
integration which uses variable time steps for efficient
computation. We solve the problem as an initial value problem.

\subsection{Initial conditions}

The structure of the governing equations makes them sensitive to
the initial values of the variables. Care should therefore be
taken to determine these values accurately. In particular, as
(\ref{37}) is singular for $\kappa^*=0$, finite initial conditions
for $u_c$ and $\sigma$ need to be identified.

Bhaumik (2005) compares three different concepts for determining
the initial conditions:

$\bullet$ Power series (McDougall, 1978)

$\bullet$ Virtual point source (Ditmars and Cederwall, 1974)

$\bullet$ Densimetric Froude number (W{\"u}est et al., 1992).

The conclusion of Bhaumik is that the method of W{\"u}est et al.
is superior to the others as it preserves the multiphase nature of
the plume through the incorporation of the phase fraction
$\alpha_g$ and relative width $\lambda$ and is independent of any
arbitrary parameters. This method is accordingly used in the
following.

As for the critical conditions on $r_b$, the main parameter
controlling the  bubble size is the pore size of the source (or
diffuser) ( W{\"u}est 1992, Bhaumik 2005). These authors point out
that the the initial bubble size depends on the gas flow rate,
larger flow rates yielding larger bubbles and vice versa. The
exact correlation between flow rates and bubble sizes is however
not known. For simplicity, we therefore assume in the following
that the conditions are ideal in the sense that the initial bubble
sizes are determined solely from the orifice diameter  if not
provided directly from observations.

Next, consider the initial condition on $\sigma$. We will follow
W{\"u}est (1992) in taking the initial plume radius to be given by
the apparent size of the bubble source area, thus considering the
entire diffuser area a a uniform source. The diffuser area $A_d$
equals the initial area covered by the {\it core},
\begin{equation}
A_d=\pi (\lambda b^0)^2, \label{39}
\end{equation}
where $b^0$ is the initial width of the top-hat distribution used
by W{\"u}est, and $\lambda$ is the relative with introduced
earlier. In our notation, the top-hat width $b$ is equal to the
standard deviation $\sigma$ (this comes from the condition that
the number and momentum fluxes of the plume should be independent
of which distribution is used). Thus $b^0=2\sigma^0$ initially,
and (\ref{39}) yields
\begin{equation}
\sigma^0=\frac{r_d}{2\lambda}, \label{40}
\end{equation}
where $r_d$ is the apparent radius of the diffusor system.

Consider finally the critical centerline velocity $u_c^0$.
Following again W{\"u}est (1992), we introduce first a densimetric
Froude number $Fr$ which in our notation reads
\begin{equation}
	Fr = \frac{u_c}{4\sqrt{\lambda\sigma g \frac{\rho_w-\rho}{\rho}} } \label{41}
\end{equation}
As the density $\rho$ can be expressed in terms of phase
fractions,
\begin{equation}
\rho=\alpha_g \rho_g+(1-\alpha_g)\rho_w, \label{42}
\end{equation}
and as the void fraction $\alpha_g$ can be related to the
volumetric flux $Q_g$ of gas,
\begin{equation}
Q_g=4\pi \sigma^2\alpha_g[u_c+u_s(1+\lambda^2)], \label{43}
\end{equation}
we obtain, using $\alpha_g \ll \alpha_l$, that
\begin{equation}
u_c^0 = 2 Fr^0\sqrt{\frac{\lambda g Q_g^0}{\pi \sigma^0 (u_c^0 + u_s^0(1+\lambda^2))}}. \label{44}
\end{equation}
As $u_s^0$ is determined from $r_b^0$ and $Q_g^0$ is determined
from experimental information, the initial condition on $u_c^0$ is
determined by $Fr^0$.

\subsection{Parameter determination}

\subsubsection*{Relative distribution width $\lambda$}
The parameter $\lambda$ determines the relative width of the velocity distribution and the number density distribution. The number density profile is necessarily bounded by the velocity distribution profile (as this defines the outer boundary of the plume). The parameter $\lambda$ should thus take a value on the interval $[0,1]$.

The literature presents several values for $\lambda$, ranging from $\lambda = 0.2$ used by Ditmars and Cederwall (1974) to $\lambda = 1$ used by Crounse (2007). W\"uest (1992) used the value $\lambda = 0.8$. The value of 0.8 is also adopted in the comparative study of existing models conducted by Bhaumik (2005). The parameter $\lambda$ is taken to be 0.8 in the present work as the work by W\"uest (1992)  and Bhaumik (2005) showed that it yields good results for dissolving plumes.

\subsubsection*{Bubble slip velocity $u_s$}
For a particle moving with a steady terminal velocity $U$ in a gravitational field, the drag force balances the difference between weight and buoyancy (Clift (2005)). It is reasonable to expect that the actual slip velocity for the bubbles is different from the terminal velocity because of screening and interactions between bubbles as shown in simulations carried out by Esmaeeli and Tryggvason (1999). However, in the present work it is assumed for simplicity that the bubble slip velocity equals the terminal velocity.

The curve fitting obtained by W\"uest (1992) for terminal velocities is used in the present work and is given by
\begin{eqnarray}
	u_s(r_b) = w_1(\frac{r_b}{\tilde{r}})^{1.357}&\mbox{~if~}& \frac{r_b}{\tilde{r}} < 7.0\cdot10^{-4}\\ 
	u_s(r_b) =  w_0&\mbox{~if~}& 7.0\cdot10^{-4} <\frac{r_b}{\tilde{r}} < 5.1\cdot10^{-3}\\
	u_s(r_b) = w_2(\frac{r_b}{\tilde{r}})^{0.547} &\mbox{~if~}&\frac{r_b}{\tilde{r}} > 5.1\cdot10^{-3}
	\label{eq:Wcorr}
\end{eqnarray}

where $w_1 = 4474$ m/s, $w_0 = 0.23$ m/s, $w_2 = 4.202$ m/s and $\tilde{r} = 1$ m. $w_0$ is the reference velocity introduced in the dimensionless formulation.

\subsubsection*{Dissolution parameters}
The Ranz-Marshall equation introduces three new parameters $K$, $c_s$ and $c_i$. As shown, the mass transfer coefficient $K$ can be determined from bubble properties, given that the molecular diffusitivity $\mathcal{D}$ is known. The molecular diffusitivity for gases in water is of order $10^{-5}$ cm$^2$/s (Perry and Green 1997) and is general dependent on viscosity ($\mu$) and temperature ($T$). As the details of these parameters would complicate the model further, the generic value of $\mathcal{D} = 10^{-5}$ cm$^2$/s is chosen.

The solubility $c_s$ is in general a complicated function of the thermodynamical state of the system. For moderate pressures (up to $\approx$ 50atm), the solubility can be expressed by Henry's law:
\begin{equation}
	c_s = H p
	\label{eq:Henry}
\end{equation}
where $H$ is Henry's constant and $p$ is the partial pressure of the phase in question. $H$ is material and temperature dependent. Suitable values for $H$ are found in comprehensive handbooks such as Lide and Fredrikse (1995).

The rate of mass-transfer from a bubble is dependent upon the concentration of dissolved species in the surrounding water $c_i$. The process of dissolution tends to increase this concentration. Assuming that the increase in concentration is of the order of the initial concentration of the ambient fluid, and that this concentration is small compared to the saturation concentration, the in-situ concentration of dissolved gas $c_i$ is negligible compared to the solubility $c_s$. 

The partial pressure of the gaseous phase is for simplicity approximated with the hydrostatic pressure in the surroundings, thus neglecting effects of surface tension in the bubble.

\subsubsection{Turbulent correlation parameter $I$}
The turbulent correlation parameter $I$ introduced is based on the mean correlations between turbulent velocity components, i.e. a Reynolds stress and is in the present work determined from experimental data. The experiments of Milgram (1983) are chosen for calibration. The experiments were carried out with air released in an unstratified environment with negligible effects of crossflow and are thus similar to the assumptions used in the present model. Four different initial flow rates Q$_0$ were used, varying from 0.0312 kg/s to 0.767 kg/s. 

\begin{figure}[htbp]
\makebox[\textwidth][c]{
		\includegraphics[width=9cm]{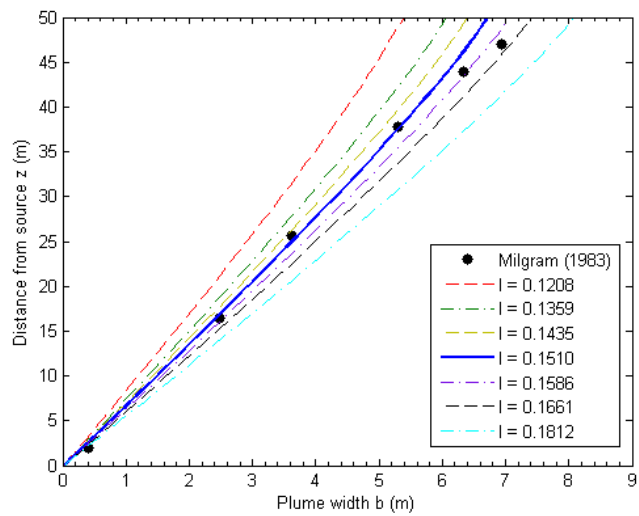}
		}
	\caption{Determination of optimal $I$ from $b$, Q$_0$ = 0.7670 kg/s. Solid line shows optimal fit for the given mass rate (based on optimum for $b$ and $u_c$), while dashed lines shows results when $I$ is chosen $\pm 5, \pm 10$ and $\pm 20 \%$ from optimum.}
	\begin{center}
	\label{fig:res1}
	\end{center}
\end{figure}

\begin{figure}[htbp]
\makebox[\textwidth][c]{
		\includegraphics[width=9cm]{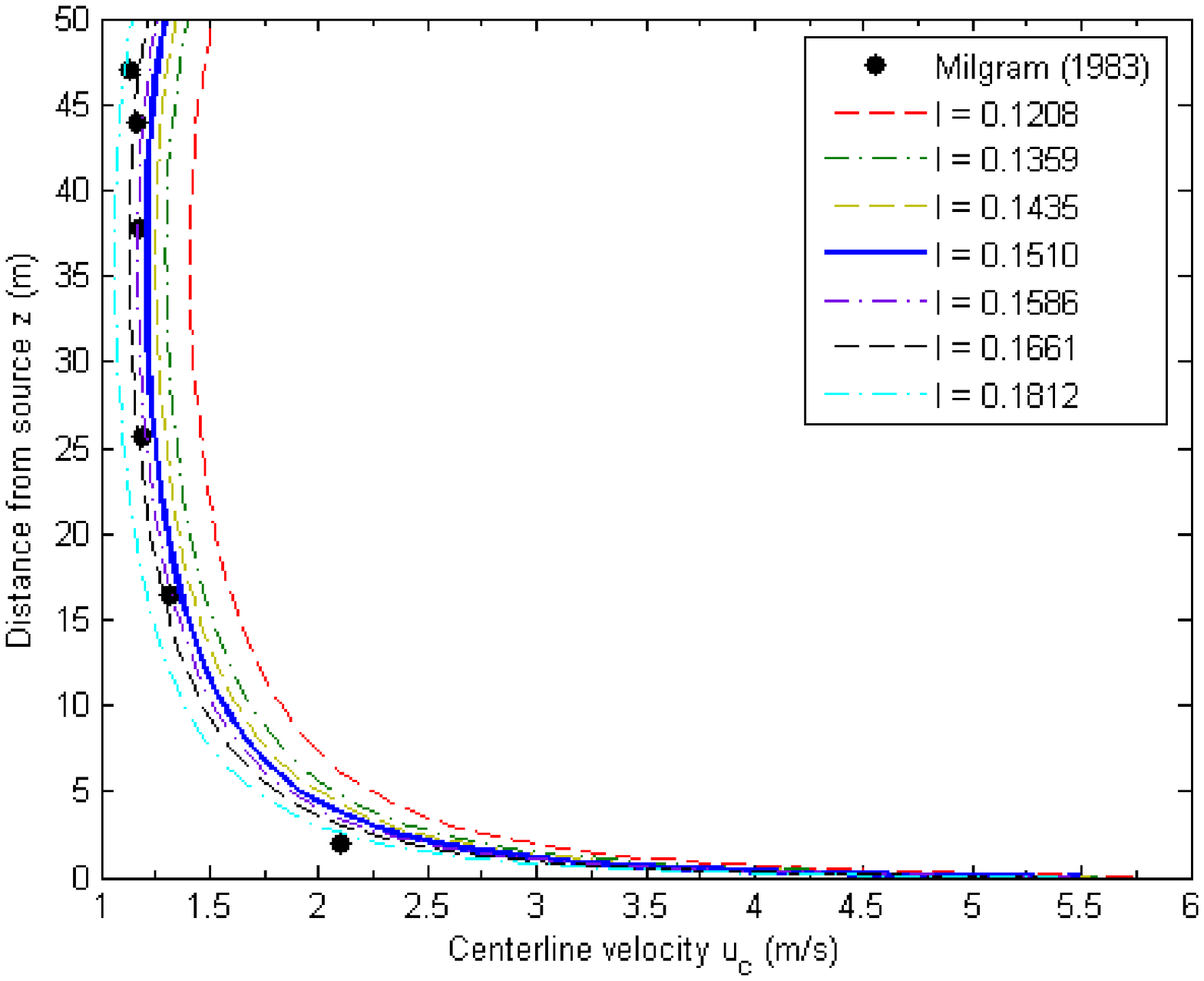}
		}
	\caption{Determination of optimal $I$ from $u_c$, Q$_0$ = 0.7670 kg/s. Solid line shows optimal fit for the given mass rate (based on optimum for $b$ and $u_c$), while dashed lines shows results when $I$ is chosen $\pm 5, \pm 10$ and $\pm 20 \%$ from optimum.}
	\begin{center}
	\label{fig:res2}
	\end{center}
\end{figure}

Using the given input data and initial conditions, the model equations are solved with $I$ as a free parameter. The parameter $I$ is varied until good accordance with the experimental data for the centerline velocity $u_c$ and plume width $b$ is obtained ($b = \sqrt{2}\sigma$). Sample plots are given in figures \ref{fig:res1} and \ref{fig:res2} and overall results are presented in tables \ref{tab:res1} and \ref{tab:res2}.

\begin{table}[!htb]
\caption{Key results from calibration from Milgram (1983). Four release rates $Q_0$ are investigated from a release depth $D$ of 50 m. The table gives relevant quantities for simulations done with optimal values for $I$ ($I_{opti}$) chosen. The mean centerline velocity $\bar{u}_c$ is calculated from the depth of release ($D$=50 m) divided by the total rise time $t_{rise}$.}
\begin{center}
\begin{tabular}{cccccc}
\hline
$Q_0$ & $Q_{surf}$ &  $b_{surf}$ & $t_{rise}$ & $\bar{u}_c$ & $I_{opti}$  \\
(kg/s) & (kg/s) & m & (s) &(m/s) & (~) \\
\hline
0.031 & 0.027 & 3.368 & 67.460	& 0.741 & 0.075 \\ 
0.153 & 0.137 & 4.598 & 47.894	& 1.044 & 0.102 \\ 
0.368 & 0.331 & 6.563 & 45.500	& 1.099 & 0.147 \\ 
0.767 & 0.703 & 6.729 & 35.973	& 1.390 & 0.151 \\ 
\hline
\end{tabular}
\end{center}
\label{tab:res1}
\end{table}

\begin{table}[!htb]
\caption{Key results from calibration from Fannel\o p and Sj\o en (1980) (Data reproduced by Milgram (1983)). Four release rates $Q_0$ are investigated from a release depth $D$ of 10 m. The table gives relevant quantities for simulations done with optimal values for $I$ ($I_{opti}$) chosen. The mean centerline velocity $\bar{u}_c$ is calculated from the depth of release ($D$=10 m) divided by the total rise time $t_{rise}$.}
\begin{center}
\begin{tabular}{cccccc}
\hline
$Q_0$ & $Q_{surf}$ &  $b_{surf}$ & $t_{rise}$ & $\bar{u}_c$ & $I_{opti}$  \\
(kg/s) & (kg/s) & m & (s) &(m/s) & (~) \\
\hline
0.007 & 0.006 & 1.002 & 12.770	& 0.783 & 0.100 \\ 
0.013 & 0.013 & 1.193 & 11.291	& 0.886 & 0.120 \\ 
0.019 & 0.019 & 1.241 & 10.085	& 0.991 & 0.125 \\ 
0.029 & 0.028 & 1.385 & 9.498	  & 1.053 & 0.140 \\ 
\hline
\end{tabular}
\end{center}
\label{tab:res2}
\end{table}

As figures \ref{fig:res1} and \ref{fig:res2} show, the model presented yields satisfactory results when compared to the experiments conducted by Milgram (1983), if the free parameter $I$ is chosen correctly. The plots do however show some deviations close to the surface. The deviations are due to plume interaction with the surface, yielding a radial jet of entrained water, described for instance by Brevik and Kristiansen (2002). Effects in the surface zone are however not considered in the present work.

Brevik and Killie (1996) point out that the turbulent correlation parameter $I$ in general could be a function of the initial mass flux $Q_0$ and depth of release $D$. Tables \ref{tab:res1} and \ref{tab:res2} show that the value of $I$ increases with increasing $Q_0$. However, even though the values of $Q_0$ in table \ref{tab:res2} are significantly lower than in table \ref{tab:res1}, the optimal values of $I$ are similar. This suggests that decreasing values of $D$ yield increasing values of $I$. 

It is not possible to combine $D$ and $Q_0$ into a dimensionless group. This suggests the presence of some other parameter also influencing the turbulent correlations. A natural choice is the viscosity coefficient $\mu$. Even though viscosity is neglected in the governing equations, it is still important for damping out turbulence at small scales. The following dimensionless group can be constructed based on the three mentioned parameters, $D$, $Q_0$ and $\mu$:
\begin{equation}	
	Re_E = \frac{Q_0}{D \mu}.
	\label{eq:ReE}
\end{equation}
The dimensionless group $Re_E$  behaves as a Reynolds number $Re$;
\begin{equation}
	Re_E = \frac{Q_0}{D \mu} = \frac{\rho u L'}{ \mu}
	\label{eq:Reyn}
\end{equation}
where $L'$ is some length scale given by the plume cross section.
In order to make predictions for an arbitrary value of $Re_E$, the values presented in tables \ref{tab:res1} and \ref{tab:res2} are fitted to a function of the form $I = a \ln{Re_E} + b$, using the least square scheme lsqnonlin in Matlab. The function is chosen because of its relatively simple form and its ability to qualitatively reproduce the behaviour observed.
The purpose of this approach is to provide a quantitative formula for predicting $I$ and to aid in achieving a qualitative physical understanding of how relevant parameters influence $I$.

\begin{figure}[ht]
\makebox[\textwidth][c]{
		\includegraphics[width=9cm]{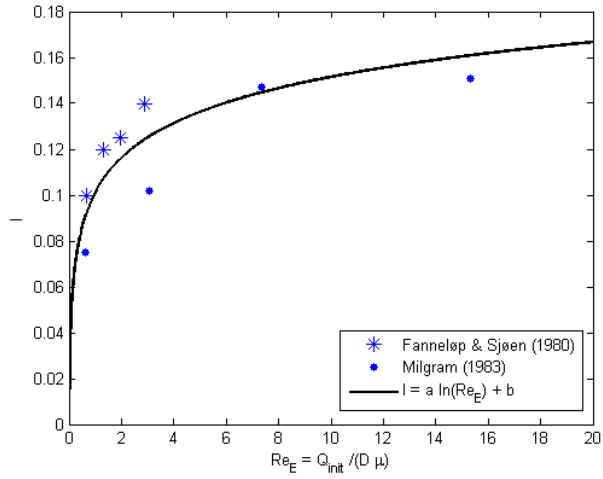}
		}
	\caption{Least square fit for relation between $I$ and $Re_E$. Solid circles show values for $I$ obtained from the experimental data of Milgram (1983) and stars show values for $I$ obtained from experimental data of Fannel\o p and Sj\o en (1980).  Solid line shows the least square fit to a function of the form $I = a\ln{Re_E}+ b$.}
	\begin{center}
	\label{fig:LSIQ}
	\end{center}
\end{figure}

The results of the least square fit is presented in figure \ref{fig:LSIQ}.

Optimal fit is obtained by choosing coefficients $a$ and $b$ corresponding to:
\begin{equation}
	I = 0.0219\ln{Re_E} + 0.1010.
	\label{eq:Iex}
\end{equation}

\section{Case studies}
In the following, a release of natural gas consisting of a mixture of 85\% methane and 15\% ethane (mass basis) will be simulated from release depths of 100 and 300 m. For both release depths mass rates of 10, 25 and 50 kg/s will be considered. The gas mixture is assumed to behave as an ideal gas and is assumed to expand isothermally. The sizes of bubbles will be varied between 1 and 10 mm and the ambient temperature and the temperature of the gas will be held constant at 278 K. The solubility of gas is calculated from Henry's law (equation \ref{eq:Henry}) and the flow is assumed to be critical at the source and the size of the source is thus determined from the given mass rate. 

The turbulent correlation parameter $I$ is determined from the initial rate of release $Q_0$ and the depth of release $D$ by equation \ref{eq:Iex}. As mentioned, equation \ref{eq:Iex} is meant as a first estimate, and experimental data exist only up to values of $Re_E \sim 20$. However, the test cases that are interesting for the industry require extrapolation well beyond this value. To deal with this, a choice is made to not allow for values of $Re_E$ predicting values of $I$ larger than 35\% of the observed values for $I$. The case of a 50 kg/s release from 100 m falls outside this limit, and is thus omitted from the analysis. Results from the case studies are presented in table \ref{tab:res3}.
\begin{table}[htb]
\caption{Key results from case studies. Mass flux at surface $Q_{surf}$, plume width at surface $b_{surf}$ and plume rise time $t_{rise}$ are computed from theory and mean centerline velocity $\bar{u}_c$ is calculated from the depth of release $D$, divided by the total rise time $t_{rise}$.}
\begin{center}
\begin{tabular}{ccccccc}
\hline
$D$ & $Q_{0}$ & $r_b^0$ & $Q_{surf}$ & $b_{surf}$ & $t_{rise}$ & $\bar{u}_c$ \\
(m) & (kg/s) & (mm) & (kg/s) & (m) & (s) & (m/s) \\
\hline
100&10&1.0&4.61&18.22&51.18&1.95\\
100&10&3.0&8.01&17.48&47.38&2.11\\
100&10&5.0&8.71&17.39&46.83&2.14\\
100&10&7.0&9.00&17.36&46.76&2.14\\
100&10&10.0&9.23&17.33&46.76&2.14\\
\hline
100&25&1.0&13.96&19.75&39.07&2.56\\
100&25&3.0&20.99&19.16&36.85&2.71\\
100&25&5.0&22.40&19.09&36.52&2.74\\
100&25&7.0&22.95&19.06&36.48&2.74\\
100&25&10.0&23.42&19.04&36.48&2.74\\
\hline
300&10&1.0& - & 0.00 & - & - \\
300&10&3.0&2.01&48.91&315.02&0.95\\
300&10&5.0&4.51&45.82&285.10&1.05\\
300&10&7.0&5.64&45.01&278.11&1.08\\
300&10&10.0&6.66&44.44&273.57&1.10\\
\hline
300&25&1.0& 0.00 & 72.78 & 404.86 & 0.74 \\
300&25&3.0&7.91&52.38&234.92&1.28\\
300&25&5.0&13.56&50.17&218.66&1.37\\
300&25&7.0&15.98&49.52&214.57&1.40\\
300&25&10.0&18.11&49.05&211.90&1.42\\
\hline
300&50&1.0&0.01&75.62&297.89&1.01\\
300&50&3.0&20.14&55.24&189.20&1.59\\
300&50&5.0&30.26&53.50&178.84&1.68\\
300&50&7.0&34.50&52.96&176.19&1.70\\
300&50&10.0&38.18&52.56&174.41&1.72\\
\hline
\end{tabular}
\end{center}
\label{tab:res3}
\end{table}
\clearpage
\section{Discussion and sensitivity analysis}
In order to investigate the robustness of the model, a sensitivity analysis of critical parameters is conducted. Sensitivity analyses for $\lambda$, $R_D$, $r_b^0$, $Fr^0$, $\mathcal{D}$ and $I$ are performed keeping other variables constant. In the sensitivity analysis the effects of parameter variation on the plume width $b$, mass-flux at the surface $Q_{surf}$ and the mean centerline velocity $\bar{u}_c$ are investigated.

The prediction of the mass-flux of the dispersed phase is relatively robust. Some variation ($\sim$ 1\%) is found when varying the bubble radius $r_b$, (smaller bubbles yielding more dissolution) and when varying the molecular diffusitivity $\mathcal{D}$ (larger diffusitivities yielding more dissolution). These findings are in accordance to those found by W\"uest (1992). The width of the plume $b$ and mean centerline velocity $\bar{u}_c$ show little sensitivity to these parameters, suggesting that the dissolution itself has little influence on the other plume properties, as long as the dissolution is moderate.

The analysis shows that the parameters of interest are insensitive to variations of the diffuser radius $R_D$ and the densimetric Froude number $Fr^0$.  The Froude number $Fr^0$ does however have some influence on the centerline velocity $u_c$ within the zone of flow establishment (ZFE), but the curves coincide further from the source. This suggests that details about the initial conditions are only important only within the ZFE and that the large scale dynamics of the plume is determined by other parameters. 

The analysis shows a strong sensitivity to the parameter $I$, an increase in $I$ of 20\% yielding an increase in the plume width of approximately 20\%. This strong dependence upon $I$ is expected and is similar to the dependence upon the entrainment coefficient $\alpha$ when the entrainment hypothesis is adopted. 

The sensitivity analysis also shows a strong dependence upon the relative distribution width $\lambda$, giving an increase in the plume width $b$ of approximately 40\% for an increase of 20\% in $\lambda$. Authors using the entrainment hypothesis have claimed that results obtained are insensitive to the relative distribution width $\lambda$ (Bhaumik (2005), Socolofsky et. al. (2002), W\"uest (1992), Milgram (1983)). The discrepancy between the well established theory in the literature and the model presented in the present work can be related to the use of the entrainment hypothesis. Use of the entrainment hypothesis allows forming a set of non dimensional differential equations which do not depend explicitly on $\lambda$, as shown for line sources by Ditmars and Cederwall (1974). Such a transformation is however not possible in the formalism based on the kinetic energy approach, giving an explicit dependence of $\lambda$ possibly explaining this discrepancy.

Even though models based on the entrainment hypothesis are relatively insensitive to $\lambda$, little agreement is found in the literature on how this parameter should be chosen, suggesting that the physics behind $\lambda$ is not clearly understood. The sensitivity analysis suggests that increasing the value of $\lambda$ yields an increasing plume width $b$ and a decreasing centerline velocity $u_c$, meaning that the more the bubbles spread out, the more efficient they are at entraining water. This suggests the possibility that $\lambda$ should \emph{not} be taken as a constant, but should depend on properties of the plume. This is an issue that ought to get some focus in future experimental or theoretical investigations.
\begin{figure}[htbp]
\makebox[\textwidth][c]{
		\includegraphics[width=9cm]{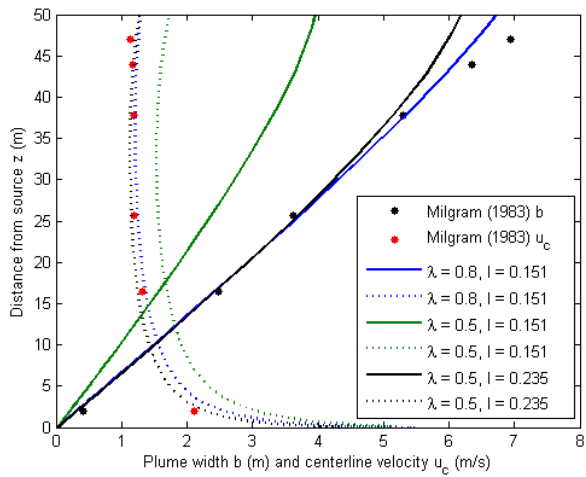}
		}
	\caption{Optimal $I$ when $\lambda$ = 0.5. Solid lines represent the plume width $b$ and dotted lines represent the centerline velocity $u_c$. Compared with experimental data of Milgram (1983) for $Q=0.7670kg/s$.}
	\begin{center}
	\label{fig:lambdaI}
	\end{center}
\end{figure}

As an alternative to the assumption that $\lambda$ is dependent on flow properties, one could investigate if the effect of choosing a different value of $\lambda$ could be incorporated in another of the free parameters of the plume.  The sensitivity analysis suggests that a change in $\lambda$ could be compensated for by a change in $I$. As shown in figure \ref{fig:lambdaI}, changing the value of $\lambda$ from 0.8 to 0.5, changes $I$ from 0.151 to 0.235 in order to obtain a satisfactory fit between theory and experiment. It is thus possible to incorporate changes in $\lambda$ in the turbulent correlation parameter $I$ and obtain, at least qualitatively, the same results. If such an approach is to be used, it should be noted that the coefficients of equation \ref{eq:Iex} should be determined for the new values of $\lambda$ as the equation only is valid under the assumption that $\lambda = 0.8$.
\section{Further issues}
\begin{description}
	\item[1] The literature contains just a few experimental data sets that are suitable for comparison with model results. This is especially the case for large rates of release, yielding poor grounds for extrapolation of the turbulent correlation parameter $I$ into these regions. Controlled laboratory experiments yield good grounds for comparison but are carried out at relatively small rates of release and small depths, making it difficult to determine the importance of dissolution. 
	 
	\item[2] The model does not distinguish between the zone of flow establishment and zone of established flow. The effects of the zone of flow establishment are believed to be handled by choosing the initial conditions accordingly. In the comparative study between existing models carried out by Bhaumik (2005) the approach using a densimetric Froude number $Fr$ introduced by W\"uest (1992) is found to be superior to its alternatives. The model used in the present work shows little sensitivity to the initial Froude number $Fr^0$ for the flow within the zone of established flow, suggesting that future work should focus on more important parameters such as the turbulent correlation parameter $I$, rather than determining exact initial conditions.
	\item[3] In the case of strong stratification the model given in the present work will not be valid as effects like detrainment are not considered. This double plume structure can be described by means of a double plume model as found in McDougall (1978) and Asaeda and Imberger (1993). These models are based on the entrainment hypothesis and yield entrainment equations for each of the plumes involved, thus introducing a higher level of complexity. An equivalent approach for the turbulent correlation parameter $I$ should be introduced if the kinetic energy approach is to advance further. 
	\item[4] The lateral variations of the turbulent stresses are assumed to be dominating over influence from vertical turbulence. As pointed out by Brevik and Kluge (1999), the desirability of taking turbulence stresses into account has been emphasized by earlier workers in the field. Brevik and Kluge (1999) model the effects of vertical turbulence by introducing a correction factor $k$. The work of these authors show that the factor $k$ is interrelated to the turbulent correlation parameter $I$. Their work indicates that no definite value of $k$ can be assigned in advance to an experimental situation without knowing details of the turbulence generating geometry of the source. A large value of $k$ ($k$ = 0.3) is introduced in order to fit the model to the experimental data of Milgram (1983). Correlations between $k$ and $I$ ought to be further investigated.
	\end{description}
\section{Concluding remarks}
In the present work, the kinetic energy approach to buoyant plumes of Brevik (1977) and Brevik and Killie (1996) is generalized in order to allow for dissolution of gas. The model is compared to experiments carried out by Milgram (1983) and is found to reproduce experimental data with satisfactory accuracy. The results presented suggest that the model presented yields a good starting point for the description of the dynamics and dissolution of gas in the zone of established flow for an air-bubble plume, given that the turbulent correlation parameter $I$ is chosen correctly.

The benefit of models based on the entrainment hypothesis is that the concept is relatively well understood and implemented for various uses. The major drawback of the approach is the difficulty of determining the value of the entrainment coefficient from parameters which are simple to measure.

Besides being relatively simple to implement, having fast convergence and promising accuracy when compared to experiments, the idea of using a conservation equation for the kinetic energy is especially interesting from a physical point of view, as it provides insight into the physics driving the plume. The mathematical framework needed is somewhat more complicated, but more physics is contained in the model. The model presented should thus be seen as an alternative to models based on the entrainment hypothesis, with the potential of answering relevant questions without excluding relevant physics. Another positive aspect of the kinetic energy approach is that the unknown turbulent correlation parameter $I$ can be determined from parameters that are easily accessible.

A combination of the simplicity of the entrainment hypothesis and the solid physical grounds of the kinetic energy approach should be sought in order to develop state of the art tools for future risk management of sub-sea gas releases, exploiting the benefits of both approaches.  

\newpage
\begin{description}
\item[~] 
Asaeda, T. and Imberger, J., 1993. Structure of bubble plumes in
linearly stratisfied environments. Journal of Fluid Mechanics 249,
35-57.
\item[~] 
Bhaumik, T., 2005. Numerical modeling of multiphase plumes: a
comparative study between two-fluid and mixed-fluid integral
models. MSc Thesis, Texas A\&M University.
\item[~]
Brevik, I., 1977. Two-dimensional air-bubble plume. Journal of the
Waterway, Port, Coastal and Ocean Division, ASCE 103, 101-115.
\item[~]
Brevik, I. and Killie, R., 1996. Phenomenological description of
the axisymmetric air-bubble plume. International Journal of
Multiphase Flow 22, 535-549.
\item[~]
Brevik, I. and Kluge, R., 1999. On the role of turbulence in the
phenomenological theory of plane and axisymmetric air-bubble
plumes. International Journal of Multiphase Flow 25, 87-108.
\item[~]
Brevik, I. and Kristiansen, {\O}., 2002. The flow in and around
air-bubble plumes. International Journal of Multiphase Flow 28,
617-634.
\item[~] 
Clift R., Grace J.~R. and Weber M.~E., 2005. Bubbles, Drops and Particles. Dover Edition, Dover Publications.
\item[~]
Crounse, B. C., Wannamaker, E. J. and Adams, E. E., 2007. Integral
model of a multiphase plume in quiescent stratification. Journal
of Hydraulic Engineering 133, 70-76.
\item[~]
Ditmars, J. D. and Cederwall, K., 1974. Analysis of air-bubble
plumes. proceedings of the 14th Conference on Coastal Engineering,
ASCE, Copenhagen, Denmark, 2209-2226.
\item[~]
Esmaeeli A.~ and Tryggvason G., 1999. Direct Numerical Simulations of Bubbly Flows Part 2. Moderate Reynolds Number Arrays, Journal of Fluid Mechanics 385, 325 - 358.
\item[~]
Fannel\o p T.~K. and Sj\o en K., 1908. Hydrodynamics of Underwater Blowouts, Proc. AIAA 18th Aerospace Sci. Meeting.
\item[~]
Johansen, {\O}., 2000. Deepblow - a Lagrangian plume model for
deep water blowouts. Spill Science \& Technology Bulletin 6,
103-111.
\item[~]
Kubasch, J. H., 2001. Bubble hydrodynamics in large pools. PhD
Thesis, Diss. ETH No. 14398, Swiss Federal Institute of
Technology, Z{\"u}rich.
\item[~] 
Lide D.~R.~ and Frederikse H.~P.~R., 1995. CRC Handbook of
Chemistry and Physics, 76th Edition. CRC Press, Inc., Boca
Raton.
\item[~]
McDougall, T. J., 1978. Bubble plumes in stratified environments.
Journal of Fluid Mechanics 85, 655-672.
\item[~]
Milgram, J. H., 1983. Mean flow in round bubble plumes. Journal of
Fluid Mechanics 133, 345-376.
\item[~]
Morton, B. R., Taylor, G. I. and Turner, J. S., 1956. Turbulent
gravitational convection from maintained and instantaneous
sources. Proceedings of the Royal Society of London, Series A,
Mathematical and Physical Sciences, 234, 1-23.
\item[~]
Perry R.~H. ~ and Green D.~W., 1997. Perry's Chemical Engineers' Handbook, 7th Edition, McGraw-Hill.
\item[~]
Ranz, W. E. and Marshall, W. R., Jr., 1952. Evaporation from
droplets, Parts I \& II. Chemical Engineering Progress 48,
173-180.
\item[~]
Socolofsky, S. A., 2001. Laboratory experiments of multi-phase
plumes in stratification and crossflow. PhD Thesis, Massachusetts
Institute of Technology.
\item[~]
Socolofsky, S. A., Crounse, B. C. and Adams, E., 2002. Multiphase
plumes in uniform, stratisfied and flowing environments.
Environmental Fluid Mechanics - Theories and Applications,
ASCE/Fluids Committee.
\item[~]
W{\"u}est, A., Brooks, N. H. and Imboden, D. M., 1992. Bubble
plume modeling for lake restoration. Water Resources Research 28,
3235-3250.
\item[~]
Zheng, L. and Yapa, P. D., 2002. Modeling of gas dissolution in
deepwater oil/gas spills. Journal of Marine Systems 31, 299-309.
\end{description}

\end{document}